# On the existence of a first order phase transition at small vacuum angle $\theta$ in the $CP^3$ model[*]

Š. Olejník[a] [‡] and G. Schierholz[b] [c]

[a]Institut für Kernphysik, Technische Universität Wien, A–1040 Vienna, Austria

[b]Deutsches Elektronen–Synchrotron DESY, D–22603 Hamburg, Germany

[c]Gruppe Theorie der Elementarteilchen, Höchstleistungsrechenzentrum HLRZ,
c/o Forschungszentrum Jülich, D–52425 Jülich, Germany

We examine the phase structure of the $CP^3$ model as a function of $\theta$ in the weak coupling regime. It is shown that the model has a first order phase transition at small $\theta$. We pay special attention to the extrapolation of the data to the infinite volume. It is found that the critical value of $\theta$ decreases towards zero as $\beta$ is taken to infinity.

## 1. INTRODUCTION

One of the characteristic features of $QCD$ is the existence of an integer valued topological charge $Q$. This gives rise to an additional term in the action,

$$S_\theta = S + i\theta Q, \ \theta \in [0, 2\pi), \tag{1}$$

where each value of $\theta$ results in a different vacuum state. A priori $\theta$ is a free parameter. Since the vacuum is an eigenstate of CP only for $\theta = 0, \pi$ and no CP violation is observed in the strong interactions, $\theta$ must, however, be very close to zero.

Is there any dynamical reason for that? There are indications that confinement at $\theta = 0$ is caused by the condensation of color magnetic monopoles [1]. In the $\theta$ vacuum these monopoles acquire an electric charge $\theta/2\pi$ [2]. One may then argue [3] that a rich phase structure will emerge as a function of $\theta$. In the weak coupling limit it is even conceivable that confinement will be limited to $\theta = 0$ only which would solve the problem rather elegantly.

Since many of the characteristic features of $QCD$ are shared by the $CP^{N-1}$ models, it suggests itself to examine these models first. We have chosen the $CP^3$ model. For smaller $N$ one may run into the problem of dislocations [4], while for large $N$ the signal one hopes to see will probably die away like $1/N$ [5].

## 2. LATTICE SIMULATION

We use the action

$$S = -2\beta \sum_{x,\mu} \text{Tr} P(x) P(x + \hat{\mu}), \tag{2}$$

where $P(x) = z(x) z^\dagger(x)$, $\text{Tr} P(x) = |z(x)|^2 = 1$, and we employ periodic boundary conditions. The action can be rewritten in terms of the complex scalar field $z(x)$ and a composite abelian gauge field interacting minimally with each other. The associated $U(1)$ link field has the form

$$U_\mu(x) = \frac{z^\dagger(x) \cdot z(x + \hat{\mu})}{|z^\dagger(x) \cdot z(x + \hat{\mu})|}, \ \mu = 0, 1. \tag{3}$$

The topological charge is obtained from (3) by forming plaquette fields $U_p = e^{iF_{01}}$, $-\pi < F_{01} \leq \pi$ and writing

$$Q = \frac{1}{2\pi} \sum_p F_{01}. \tag{4}$$

As can easily be seen, $Q$ is an integer.

In this talk we will restrict ourselves to the calculation of the partition function and a few quantities which derive from it. A full account of our work will be given elsewhere [6]. The partition function reads

$$Z(\theta) = \sum_Q e^{i\theta Q} p(Q), \tag{5}$$

---

[*]Talk presented by G. Schierholz.
[‡]On leave from Institute of Physics, Slovak Academy of Sciences, SK–84228 Bratislava, Slovakia.

$$p(Q) = \frac{\int [\mathcal{D}z\mathcal{D}z^\dagger]_Q \delta(|z|^2 - 1)\mathrm{e}^{-S}}{\int \mathcal{D}z\mathcal{D}z^\dagger \delta(|z|^2 - 1)\mathrm{e}^{-S}}, \qquad (6)$$

where the subscript $Q$ indicates that the path integration is restricted to the given charge sector. In order to compute $p(Q)$ we proceed as follows [7]. We divide the phase space into overlapping sets of five consecutive charges. In each of these sets the $z$ fields are updated by a combination of Metropolis and overrelaxation steps, $z' = \exp(\mathrm{i}\phi_a \lambda_a)z$, where the $U(4)$ generators $\lambda_a$ are selected randomly [8]. This is supplemented by an additional acceptance criterion: if the new charge is in the same set the configuration is accepted and the new charge recorded; if the new charge is outside the set the change is rejected and the old charge recorded. Furthermore, a trial charge distribution which is approximately equal to the true distribution is incorporated.

So far we have done simulations at two values of $\beta$ on various lattice volumes. We have chosen $\beta = 2.5$, $V = 28^2, 32^2, 38^2, 48^2$, and $\beta = 2.7$, $V = 46^2, 56^2, 64^2, 72^2$. The correlation lengths at these two $\beta$ values are [9] $\xi \approx 4.5$ and $\xi \approx 8.8$, respectively. On these lattices we are able to follow $p(Q)$ over more than twenty orders of magnitude.

## 3. RESULTS

The first quantity we looked at is the free energy $F(\theta)$ which is given by

$$Z(\theta) = \mathrm{e}^{-V F(\theta)}. \qquad (7)$$

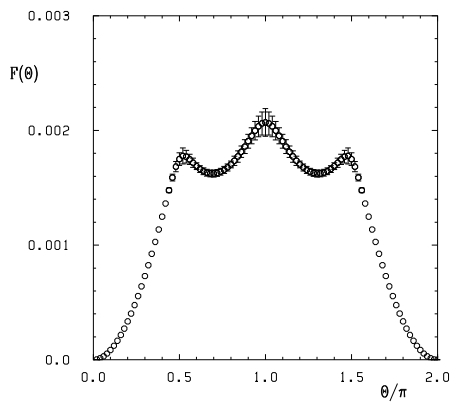

Figure 1. The free energy $F(\theta)$.

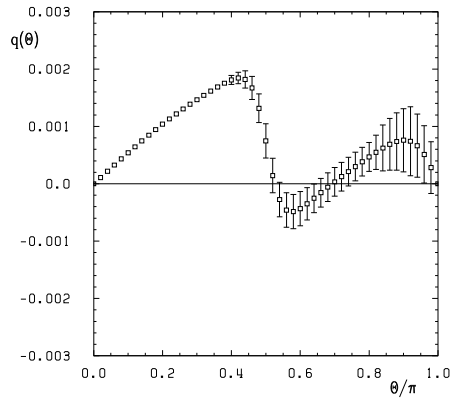

Figure 2. The average topological charge $q(\theta)$. In the second half of the $\theta$ interval not shown here $q(\theta) = -q(2\pi - \theta)$.

A first order phase transition will manifest itself in a kink in $F(\theta)$. In Fig. 1 we show $F(\theta)$ on the $V = 64^2$ lattice at $\beta = 2.7$. The error bars have been computed by a jackknife method. We see two kinks, one at $\theta \approx 0.5\pi$ (and a corresponding one at $\theta \approx 1.5\pi$) and one at $\theta = \pi$. This should be compared with the prediction of the dilute instanton gas approximation, $F(\theta) \propto 1 - \cos\theta$, and the prediction of the $1/N$ expansion [5], $F(\theta) \propto \theta^2$. Clearly, our results do not agree with either of the two. In the following we shall mainly be interested in the phase transition at small $\theta$.

The first derivative of $F(\theta)$ is equal to the average topological charge $q(\theta)$:

$$\frac{\mathrm{d}F(\theta)}{\mathrm{d}\theta} \equiv q(\theta) = -\frac{\mathrm{i}}{V}\frac{1}{Z(\theta)}\sum_Q \mathrm{e}^{\mathrm{i}\theta Q} Q\, p(Q). \qquad (8)$$

At a first order phase transition this quantity must be discontinuous. In Fig. 2 we show $q(\theta)$ for the $V = 64^2$ lattice at $\beta = 2.7$. We see that $q(\theta)$ increases almost linearly with $\theta$ until it reaches the location of the phase transition where it jumps to a value which is close to zero. According to (4), $q(\theta)$ can be interpreted as a background electric field. In Ref. [6] we shall argue that at the phase transition the charged scalars (associated with the field $z(x)$) are liberated which causes the background electric field to collapse.

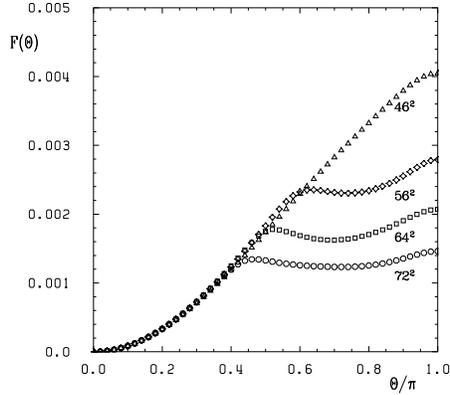 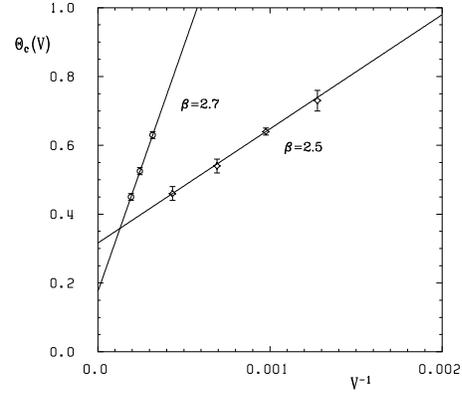

Figure 3. The free energy $F(\theta)$. The error bars are omitted in order to keep the figure legible. They are comparable to the ones in Fig. 1.

Figure 4. The critical value of $\theta$.

Let us now look at the volume dependence of the free energy. In Fig. 3 we show $F(\theta)$ for four different volumes at $\beta = 2.7$. We see that the location of the phase transition moves to smaller values of $\theta$ as $V$ is increased. Below the phase transition all data points fall on a universal curve. If the lattice volume is too small we see no signal of a phase transition [10].

We denote the critical value of $\theta$ by $\theta_c$ and identify it with the local minimum of $Z(\theta)$. The volume dependence of $\theta_c$ is linked to the order of the phase transition. For a first order transition we expect

$$\theta_c(V) - \theta_c(\infty) \propto V^{-1}. \qquad (9)$$

In Fig. 4 we show $\theta_c$ as a function of $V^{-1}$ for $\beta$ = 2.5 and 2.7. In both cases our data fall on a straight line in accordance with a first order phase transition. This allows us to extrapolate the data to the infinite volume.

We obtain $\theta_c(\infty) = 0.32(2)$ at $\beta = 2.5$ and $\theta_c(\infty) = 0.18(3)$ at $\beta = 2.7$. It is perhaps not surprising that the critical value of $\theta$ depends strongly on $\beta$. Note that $\theta_c = \pi$ in the strong coupling limit [11].

## 4. CONCLUSIONS

We have found a novel, first order phase transition in $\theta$ with $\theta_c(\infty)$ decreasing towards zero as we approach the continuum limit. This transition is consistent with a deconfining phase transition [6].

If somebody wants to repeat the calculation he should start with $\beta = 2.7, V = 56^2$.